\documentclass[12pt,epsf]{article}

 \usepackage{epsfig}
 \usepackage{times}
 \usepackage{epstopdf}
 \usepackage{amsmath}
 \usepackage{amssymb}

\newcommand{\be}{\begin{equation}}
\newcommand{\ee}{\end{equation}}
\def\({\left (}
\def\){\right )}
\def\[{\left [}
\def\[{\right ]}

\begin{document}
\begin{titlepage}
\bigskip
\rightline
\bigskip\bigskip\bigskip\bigskip
\centerline {\Large \bf {New instabilities of de Sitter spacetimes}}
\bigskip\bigskip
\bigskip\bigskip

\centerline{\large Keith Copsey and Robert Mann}
\bigskip\bigskip
\centerline{\em Department of Physics, University of Waterloo, Waterloo, Ontario N2L 3G1, Canada}
\centerline{\em kcopsey@scimail.uwaterloo.ca, mann@avatar.uwaterloo.ca}
\bigskip\bigskip

\begin{abstract}

We construct an instanton describing the pair production of non-Kaluza Klein bubbles of nothing in higher odd dimensional de Sitter spaces.  In addition to showing that higher dimensional de Sitter spaces have a nonzero probability to become topologically nontrivial, this process provides direct evidence for the association of entropy with cosmological horizons and that non-Kaluza Klein bubbles of nothing are a necessary ingredient in string theory or any other consistent quantum theory of gravity in higher dimensions.

\end{abstract}
\end{titlepage}

\baselineskip=16pt
\setcounter{equation}{0}

\section{Introduction}

Black hole pair production is a well-established process in semi-classical quantum gravity in
which pairs of black holes are
created via a  Schwinger-like process from an external field.  The masses of the black holes
produced  can take on any value consistent with standard conservation laws.   The energy needed to create the black holes comes from the energy of the background field, which also has provides the
force necessary to accelerate the black holes once they are created.  Background fields
such as an external electromagnetic field with its Lorentz force \cite{EMpair}, a cosmological constant
(or inflation) \cite{RossMannBH,cosmo}, a cosmic string \cite{cstring}, and a domain wall \cite{dwall} (and various combinations \cite{combo}, including rotation \cite{Ivan})  have all been shown to generate this process in 4 spacetime dimensions.

These studies have repeatedly provided us
with evidence that the exponential of the entropy of a black hole does indeed correspond
to the number of its quantum states.  Although the instability of de Sitter spacetime has a long history \cite{Ginsparg}, the role of the cosmological horizon has been less than
clear in this regard, with alternate arguments being employed \cite{Bousso} to suggest a similar
situation holds in this case. 

Here we demonstrate that higher-dimensional de Sitter spacetime has a new kind of instability
to decay into soliton pairs (or pairs of bubbles of nothing).   String theories find their firmest footing in higher-dimensional contexts and charged black hole pair production has been shown to take place in
this context as well \cite{DiasLemos}.  We find that a similar mechanism generating soliton pairs
 provides direct evidence that one should associate an entropy with a cosmological horizon that is
a quarter of its area.  This pair production process will compete with that of black hole pair production in dimensions $d\geq 5$ and, in fact, exceed it if the black hole radius is sufficiently large in five dimensions.   Higher dimensional de Sitter spaces therefore have a nonzero probability to decay into a topologically nontrivial  spacetime.

\section{de Sitter Bubbles}

Here we outline the construction of the $d\geq 5$  de Sitter solitons that we consider as decay products of de Sitter spacetime.

\subsection{de Sitter solitons}

Beginning with the bulk action\footnote{We will consider specific surface terms later as necessary.}
\be \label{act1}
S = \frac{1}{16 \pi G_d} \int \sqrt{-g} (R - 2 \Lambda)
\ee
we wish to find solutions for $\Lambda > 0$ besides the well-known black hole solutions.
Using the fact that one may write an odd dimensional (round) sphere $S^{2N + 1}$ as an $S_1$ fibered over $CP^{N}$, consider the following ansatz for $d$-dimensional de Sitter spacetimes containing a squashed sphere
\be \label{metric1}
ds^2 = -g(r) dt^2 + \frac{dr^2}{f(r) g(r)} + r^2 f(r) (d\chi + A)^2 + r^2 d \Sigma^2
\ee
where $\chi$ is a periodic direction with period $2 \pi$, $d \Sigma^2$ is the metric for  $CP^{(d-1)/2}$ and $A$ the usual  one form on the $CP$ base space.  Written this way, the metric on the round sphere is
\be
d \Omega_{d-2} = (d \chi + A)^2 + d \Sigma^2
\ee
For the sake of convenience, we review this fibration in detail and give explicit forms for the first several cases in an appendix.  Given the ansatz (\ref{metric1}),  $f(r)$ and $g(r)$ may be solved for uniquely as\footnote{To be precise we have checked this solutions for five, seven, nine, and eleven dimensions, although it appears fairly clear the same solutions work for any odd dimensions larger than three.}
\be \label{gftn}
g(r) = -\frac{r^2}{l^2} + 1
\ee
and
\be \label{fftn}
f(r) = 1 - \frac{r_0^{d- 1}}{r^{d - 1}}
\ee
where $l$ is the usual de-Sitter length, i.e.
\be
\Lambda = \frac{(d-2) (d-1)}{2 l^2} 
\ee
There are more generic solutions than (\ref{metric1}) one might consider, in particular by allowing $g_{r r}$ to be a generic function of $r$.  However, the standard asymptotics do not appear to allow any analytic solutions besides those given above.  One could also consider adding additional matter, but for the present we will allow only a cosmological constant.

Near $r = r_0$ the direction $\partial/{\partial \chi}$ degenerates, and one is left with a minimal compact surface, sometimes known as a bubble of nothing, formed in this case from the $CP^{N}$.  In the five dimensional case, $CP^1 = S^2$ and the solution reduces to that found by \cite{ClarksonMann1} and later discovered independently as an example of a much wider class of time symmetric initial data in \cite{CopseyBubblesII}.  A broad class of higher-dimensional versions was pointed out in \cite {MannClarksonII}.

Demanding the absence of a conical singularity at the bubble (at $r = r_0$), up to a $Z_k$ orbifold, implies
\be
r_0 = l \sqrt{1 - \frac{4}{(d-1)^2 k^2}}
\ee
Intuitively, de Sitter space tends to make things expand and hence a positive cosmological constant allows for a stationary solution.  If one writes down the same solution for zero or negative cosmological constant one is either forced to allow a conical singularity at the bubble or to quotient the spacetime, reflecting the fact that without some additional force gravity tends to make the bubble collapse.  Topologically these solutions are simply connected\footnote{$\pi_2$ is, however, nontrivial.}
 and one may argue, unlike Kaluza-Klein bubbles \cite{Wittenbubble}, there is no obstruction to defining a spin structure on such solutions \cite{CSsol}.   Since we will describe below an instanton nucleating such bubbles from empty de Sitter space, the latter point is, of course, not surprising.

\subsection{de Sitter instantons}

Now consider the Euclidean solution obtained by continuing $t \rightarrow i \tau$.  Since $\partial/{\partial \tau}$ degenerates at $r =l$, the cosmological horizon, we are forced to periodically identify $\tau$.   Demanding the absence of a conical singularity at $r = l$ implies $\tau$ has period
\be
\beta = \frac{2 \pi l}{\sqrt{1 - \frac{r_0^{d-1}}{l^{d-1}}}}
\ee
Note that unlike the more familiar black hole solutions, there is no second place where $\partial/{\partial \tau}$ degenerates and one need not set two temperatures equal here.  The bubble has no entropy of its own but merely acts like a mirror, reflecting the incoming thermal radiation back outwards.

We have obtained a compact Euclidean solution, since $\tau$ is compact and $r$ ranges over a finite range (specifically from $r_0$ to $l$).  Slicing this solution along a moment of time symmetry yields a compact instanton describing the production of the above Lorentzian solution.  Properly speaking, the instanton describes the pair production of solitons, since the static patch only covers part of de Sitter space (see, e.g., \cite{GHdeSitter}).   To calculate the rate of bubble nucleation from this instanton we require the on-shell Euclidean action.  Since there is no matter present, besides $\Lambda$, the only surface terms that are needed to be added to give a good variational principle to the action (\ref{act1}) are gravitational.  As we wish to find the instanton ending on a time symmetric surface $\Sigma$ with the given spatial metric (i.e. the $t = constant$ slice of (\ref{metric1})), we are solving a gravitational Dirichlet problem with the metric defined exactly on a given surface.  Hence the appropriate surface term is the famous Gibbons-Hawking term \cite{GHbdy}.  However, since $\Sigma$ is a surface of time symmetry, its extrinsic curvature vanishes and we simply obtain the on-shell action for the instanton
\be
S_E = - \frac{1}{16 \pi G_d} \int d^d x \sqrt{g} [ R - 2 \Lambda ] = - \frac{l^{d-2} \Omega_{d - 2}}{8 G_d} \sqrt{1 - \frac{r_0^{d-1}}{l^{d-1}}}
\ee
Note this is the action for the instanton nucleating bubbles, not the action for a bounce, although in this case the two are related by a factor of two.

Following the standard Coleman-de Luccia prescription \cite{CdL}, the probability to nucleate such bubbles is
\be
P_{bubble} \approx e^{-2 S_E + S_0}
\ee
where $S_0$ is the Euclidean action of the false vacuum, namely pure de Sitter space
\be
S_0 = -\frac{l^{d -2} \Omega_{d -2}}{4 G_d}
\ee
Then
\be
P_{bubble} \approx  \mathrm{exp}\Bigg[-\frac{ 2 \Omega_{d -2}}{(2 \pi)^{d- 4}} \frac{l^{d-2}}{l_p^{d-2}}  \Bigg( 1-\sqrt{1 - \frac{r_0^{d-1}}{l^{d - 1}}}  \Bigg) \Bigg]
\ee
where we have related the $d$-dimensional Newton constant to the $d$-dimensional Planck length
\be
16 \pi G_{d} = (2 \pi)^{d -3} l_p^{d - 2}
\ee
The timescale for this process to take place is
\be \label{tdec}
\tau_{decay} \approx e^{2 S_E - S_0} \approx e^{2 S_E} \tau_R
\ee
where $\tau_R$ is the recursion time for pure de Sitter space.  Since $S_E$ is negative and large, in Planck units, whenever one trusts the instanton approximation, $\tau_{decay} \ll \tau_R$ and we need not concern ourselves with well-known quantum mechanical objections to eternal de Sitter space \cite{conceptual,MSS,Susskind}.

Note that area of the cosmological horizon at $r = l$
\be
A_H = l^{d - 2} \Omega_{d - 2} \sqrt{1 - \frac{r_0^{d-2}}{l^{d-1}}}
\ee
is reduced from the area of the pure de Sitter space that we began with.   Then, presuming one believes that cosmological horizons have an entropy proportional to their areas, the nucleation of a bubble results in a decrease in entropy.  At first glance, one might think the above was a contradiction of the second law of thermodynamics.   The second law, of course, may be violated by rare statistical fluctuations.   The time scale for pure de Sitter to fluctuate into any particular state is given by the recursion time scale
\be
\tau_R \approx e^{-S_0}
\ee
On the other hand, if one asks the time scale for de Sitter to fluctuate not into a particular state but into any one of N states in an ensemble,
\be \label{tN}
\tau_N  \approx \frac{\tau_R}{N}
\ee
It is this second comparison that is relevant here--we are considering a de Sitter-bubble at finite temperature, specifically at
\be
T = \frac{1}{k_B \beta} = \frac{1}{k_B 2 \pi l} \sqrt{1 - \frac{r_0^{d-1}}{l^{d-1}}}
\ee
Comparing (\ref{tdec}) to (\ref{tN}), this then implies
\be
N = e^{-2 S_E} = e^{\frac{A_H}{4 G_d}}
\ee
That is, we are considering a state with entropy $A_H/{4 G_d}$.   Hence the above instanton provides direct evidence that one should associate an entropy to a quarter the cosmological horizon size.  Note this argument is far more direct and compelling than the analogous argument using black hole pair production, for in the second case one must, as far as we know, simply assert that any black hole  has an entropy a quarter of its horizon area, independent of whether its asymptotics are de Sitter, anti de Sitter, or asymptotically flat.

It is worth noting there are other de Sitter bubble-nucleating instantons one can consider if one does not mind enlarging the set of asymptotics under consideration.   In particular, one can consider the solutions of \cite{MannClarksonII}
$$
ds^2 = -g(r) dt^2 + \Big( \frac{2 r}{d-1} \Big)^2 f(r) \Big[ d \chi + \Sigma_{i = 1}^{k} \cos \theta_i \, d\phi_i \Big]^2
$$
\be \label{squashedmetric}
+ \frac{dr^2}{g(r) f(r)} + \frac{r^2}{d-1} + \Sigma_{i = 1}^{k} (d\theta_i^2+ \sin^2 \theta_i \, d\phi_i^2)
\ee
where $k = (d-3)/2$ and $g(r)$ and $f(r)$ are the same as in (\ref{gftn}) and (\ref{fftn}).  Asymptotically (\ref{squashedmetric}) may be described as containing squashed spheres--it is straightforward to check that the square of the Riemann tensor for the angular section of (\ref{squashedmetric}) (i.e. $r$ and $t$ held constant) is different from that of round spheres.   Nevertheless, one may calculate an instanton as above and  obtain  qualitatively similar results.

\setcounter{equation}{0}
\section{Comparison with black hole production}

We wish to compare the relative rate of production of the above bubbles of nothing to another well known instability in de Sitter space--the nucleation of charged pairs of de Sitter black holes \cite{RossMannBH, DiasLemos}.  For the sake of simplicitly, we will limit ourselves to electrically charged non-extremal solutions--the so-called ``lukewarm'' solutions where one sets the temperature of the black hole to match that of the cosmological horizon.  The consideration of charge is a necessary complication--one is unable to match the black hole and cosmological horizon temperatures without some form of matter.   We consider the bulk action
\be
S_B = \frac{1}{16 \pi G_d}  \int_{\mathcal{M}} \sqrt{g} (R - 2 \Lambda - F_{a b} F^{a b})
\ee
again omitting surface terms until we need them later.   The black hole solutions may be written as
\be \label{BHsol}
ds^2 = -f(r) dt^2 + \frac{dr^2}{f(r)} + r^2 d\Omega_{d - 2}
\ee
where $d \Omega_{d - 2}$ is the metric on the unit $d-2$-sphere and
\be \label{fdef}
f(r) = -\frac{r^2}{l^2} + 1 - M r^{3-d} + Q_1^2 r^{6 - 2 d}
\ee
and the field strength is
\be
F_{r t} = Q_0 r^{2 - d}
\ee
where
\be
Q_0 = \sqrt{\frac{ (d - 2) (d- 3)}{2}} Q_1
\ee

The prescription of \cite{RossMannBH} is that the appropriate analytic continuation is $t \rightarrow i \tau$ leaving $Q_1$ real, that is to allow a complex field strength.  The sensbility of this prescription is argued in \cite{RossMannBH}, although perhaps all other concerns are trumped by the fact that, at least in four dimensions, one can not find a smooth instanton if one continues $Q_0 \rightarrow i Q_0$ since  the temperatures of the black hole and cosmological horizon can not be matched \cite{Ivan}.   Defining the largest (real) zero of f as $r = r_c$ and the second largest as $r = r_+$, that is the cosmological and outer black hole horizons respectively, one quickly finds the period of Euclidean time must be
\be
\beta = \frac{4 \pi}{f'(r_+)} = \frac{4 \pi}{\vert f'(r_c) \vert}
\ee
The topology of the Euclidean solution is $S^2 \times S^{d-2}$, where the first $S^2$ is parametrized by $(r, \tau)$; note here $\partial/{\partial \tau}$ degenerates at both the cosmological and the black hole horizon.   To find the instanton we cut the full Euclidean solution at a moment of time symmetry, leaving the instanton with a boundary of topology $S^1 \times S^{d - 2}$.  In terms of the above coordinate system the relevant $S_1$ may be written as the union of the $\tau = 0$ and $\tau =\beta/2$ surfaces.

The Euclidean action is given here for the manifold $\mathcal{M}$ with boundary $\mathcal{\delta M}$ by
$$
S_E = -\frac{1}{16 \pi G_d}  \int_{\mathcal{M}} \sqrt{g} (R - 2 \Lambda - F_{a b} F^{a b})
$$
\be \label{BHact}
 - \frac{1}{8 \pi G_d} \int_{\mathcal{\delta M}}  \sqrt{h} K - \frac{1}{4 \pi G_d } \int_{\mathcal{\delta M}} \sqrt{h} n_a F^{a b} A_b 
\ee
where $K$ is the extrinsic curvature of, and $n_a$ the unit normal to, $\mathcal{\delta M}$.  The fact that these are the appropriate surface terms requires a bit of explanation.  The Gibbons-Hawking term is appropriate since it yields a well-defined variational principle provided one specifies the metric exactly on some surface.  Here $\mathcal{\delta M}$ is a compact surface and we require the metric on it to match that of the instanton we are constructing.  However, since we slice the instanton at a moment of time symmetry, $K = 0$ and this term vanishes, just as was the case for the bubble instanton above.

For the field $F$ one must specifiy either the potential $A_b$ or the normal component of the field strength $n_a F^{a b}$ on $\delta M$.  Note the potential we are talking about here is not just the potential at the cosmological horizon or infinity but throughout the bulk; on physical grounds there should still be gauge freedom in the bulk so we fix the normal component of the field strength to match the instanton.   Given this boundary condition, the second surface term in (\ref{BHact}) yields a good variational principle for the field $F$.   One can also argue \cite{RossHawkingDuality} that in four dimensions electromagnetic duality forces this choice provided one regards the magnetic charge as fixed\footnote{If one does not fix the magnetic charge, the value of the Hamiltonian will not be fixed (see, e.g., \cite{CopseyHorowitz1st})}.   At first glance this term does not appear gauge invariant but provided $\mathcal{\delta M}$ is either compact, as it is in the present case, or has boundaries upon which the potential is specified (e.g. at infinity) this is just an illusion and the surface term and the action are gauge invariant.  The only exception to the above is if one chose a gauge which corresponds to a singular field strength.  In fact if one tried to take a simple time independent potential for $A$
\be
A_\tau = \frac{i Q_0 }{3-d} r^{3-d} + C_0
\ee
one can not chose the constant $C_0$ so that $A_\tau$ vanishes at both the black hole and cosmological horizon.   The failure of the potential to vanish at the points where $\tau$ degenerates corresponds to a diverging potential and a $\delta$-function field strength (see, e.g., \cite{CSsol} for details).   Instead one may take a gauge
\be
A_r = -i Q_0 \tau r^{2 - d}
\ee
This potential might appear to be discontinuous at the horizons (where the $\tau = 0$ and $\tau = \beta/2$ surfaces meet) but going to a set of orthonormal coordinates it is easy to see the physical potential vanishes at these points\cite{RossMannBH}.

Given all the above
\be \label{GenBHact}
S_E = -\frac{\beta \, \Omega_{d-2}}{16 \pi G_d} \Big[\frac{r_c^{d-1} - r_+^{d-1}}{l^2} + (d-3) Q_1^2 (r_c^{3-d} - r_+^{3-d}) \Big]
\ee
where 
\be
\Omega_{d - 2} = \frac{2 \pi^{(d-1)/2}}{\Gamma \Big( \frac{d-1}{2} \Big)}
\ee
is the usual area of the unit $(d-2)$-sphere.  If one tried directly to write$\beta$,  $r_+$ and $r_c$ in terms of various physical parameters ($Q_1, l, M, \ldots$) one runs into the complications of roots of high order polynomials.  Fortunately this may avoided as follows.  It is useful to parametrize the ratio between the black hole outer horizon and cosmological horizon as
\be
x = \frac{r_+}{r_c}
\ee
and so $0 < x < 1$.  Let us henceforth restrict our attention to odd dimensions, since in even dimensions at present we have no bubbles to compare to.\footnote{We have checked all the various technical results below for general $d$ up to eleven dimensions, although, as before, it seems clear analogous results are valid generically.}   In odd dimensions we may write
$$
f(r) =  -\frac{r^2}{l^2} + 1 - M r^{3-d} + Q_1^2 r^{6 - 2 d} 
$$
\be \label{fexp}
= -\frac{r^2}{l^2} \Big( 1 - \frac{r_c^2}{r^2} \Big) \Big(1 - \frac{ x^2 r_c^2}{r^2} \Big) \Big( 1 + \Sigma_{n = 1}^{d-4} a_n \frac{r_c^{2 n}}{r^{2 n}}\Big)
\ee
The absence of a conical singularity
\be
f'(x r_c) + f'(r_c) = 0
\ee
in this parametrization is equivalent to
\be \label{con2}
1+\Sigma_{n = 1}^{d - 4} \frac{a_n}{x^{2 n}} = x (1 + \Sigma_{n = 1}^{d-4} a_n )
\ee
One may then solve (\ref{fexp}) for $M$, $Q_1^2$ and the $a_n$'s in terms of $r_c$, $x$, and $l$.   The absence of a conical singularity (\ref{con2}) then fixes $r_c$ in terms of $x$ and $l$.  For five dimensions this gives
\be
a_1 = -\frac{x^2}{1+x+x^2}
\ee
\be
M = \frac{ l^2 x^2 (1 + x + x^2) (2 + x + 2 x^2)}{(1+x+3 x^2 + x^3 + x^4)^2}
\ee
\be
Q_1^2 = \frac{l^4 x^4 (1 + x + x^2)^2}{(1 + x + 3 x^2 + x^3 + x^4)^3}
\ee
and
\be
r_c^2 = \frac{l^2 (1 + x + x^2)}{1 + x + 3 x^2 + x^3 + x^4}
\ee
For higher dimensions one obtains similar, but increasingly complex and unilluminating expressions.  Once one finds all the above constants in terms of $x$ and $l$ one can show the euclidean action (\ref{GenBHact})  can be written in the remarkably simple
\be
S_E = -\frac{\Omega_{d -2}}{8 G_d} (r_+^{d-2} + r_c^{d-2})
\ee
In fact, just as for the bubbles, the fact that the action takes this form is crucial to avoiding a contradiction with the second law of thermodynamics.

We note that, recalling that $0 < x < 1$,
\be
r_c < l
\ee
or in other words the cosmological horizon is smaller in the presence of the black holes than in empty de Sitter, just as is true for the bubbles.  Further the sum of the horizon areas is less than the size of the cosmological horizon of empty de Sitter space
\be
A_H = \Omega_{d-2} (r_+^{d-2} + r_c^{d-2}) < \Omega_{d -2} l^{d-2}
\ee
Hence, as with the bubbles, one might have feared one is violating the second law of thermodynamics.  
As in the case of the bubbles this process may be described as the expected statistical fluctation provided one traces over a number of states
\be
N = e^{-2 S_E} = e^{A_{\mathrm{horizons}}/{4 G_d}}
\ee

We are finally in a position to compare the action for bubble production to black hole nucleation.  It is useful to define the ratio of these probabilities in terms of a quantity $\delta$ to isolate the common dependence on $l/l_p$:
\be \label{Prat}
\frac{P^{BH}}{P^{Bubble}} \approx \frac{e^{-2 S^{BH} + S_0}}{e^{-2 S^{Bubble} + S_0}} = e^{\delta_d \frac{l^{d-2}}{{l_p}^{d-2}}}
\ee
Then one finds 
\be \label{deltad}
\delta_d = 2^{6-d} \pi^{(7-d)/2} \Bigg[ \, \Big(\frac{r_c}{l} \Big)^{d-2} (1 + x^{d-2}) - \sqrt{1 - \frac{r_0^{d-1}}{l^{d-1}}} \, \, \Bigg]
\ee
For general dimensions this seems to be the simplest expression for $\delta_d$, although in five dimensions the explicit form may be written nearly as simply
\be
\delta_5 = 2 \pi \Bigg[  (1 + x^3) \Big(\frac{1+x+x^2}{1+x+3 x^2 + x^3 + x^4}\Big)^{3/2}-\frac{\sqrt{8 k^2 - 1}}{4 k^2} \Bigg]
\ee
As $k$ increases, $r_0$ increases and $\delta$ becomes more positive.  While the $x$-dependence is not entirely obvious from (\ref{deltad}), on physical grounds one might expect small black holes to be dominant and as we show shortly the plots of $\delta$ bear out this expectation.  The numerical values of $\delta_d$ fall off rather quickly as $d$ increases.  Specifically for $k = 1$, $\delta_5 (x = 0) \approx 2.1273$, $\delta_7 (x = 0) \approx 0.22721$, $\delta_9 (x = 0) \approx 2.0810 \times 10^{-2}$, and $\delta_{11} (x = 0) \approx 1.8058 \times 10^{-3}$.  If one trusted the calculation for $l \sim l_p$, then in large dimensions the probabilities for black hole and bubble production would be always comparable.  However, since the ratio of probabilities (\ref{Prat}) also depends on $(l/l_p)^{d-2}$, (\ref{Prat}) quickly becomes either very large or very small (depending on the sign of $\delta_d$) as $l/l_p$ increases.  In order to be able to plot the various $\delta_d$ on a single graph, in Figure 1 we have plotted $\delta_d$ normalized by their values at $x = 0$. 
		\begin{figure}[t]
	\begin{picture} (0,0)
    	\put(9,8){$\frac{\delta_d}{\delta_d (0)}$}
         \put(386, -178){$x$}
    \end{picture}

\centering
	\includegraphics[scale= 0.9]{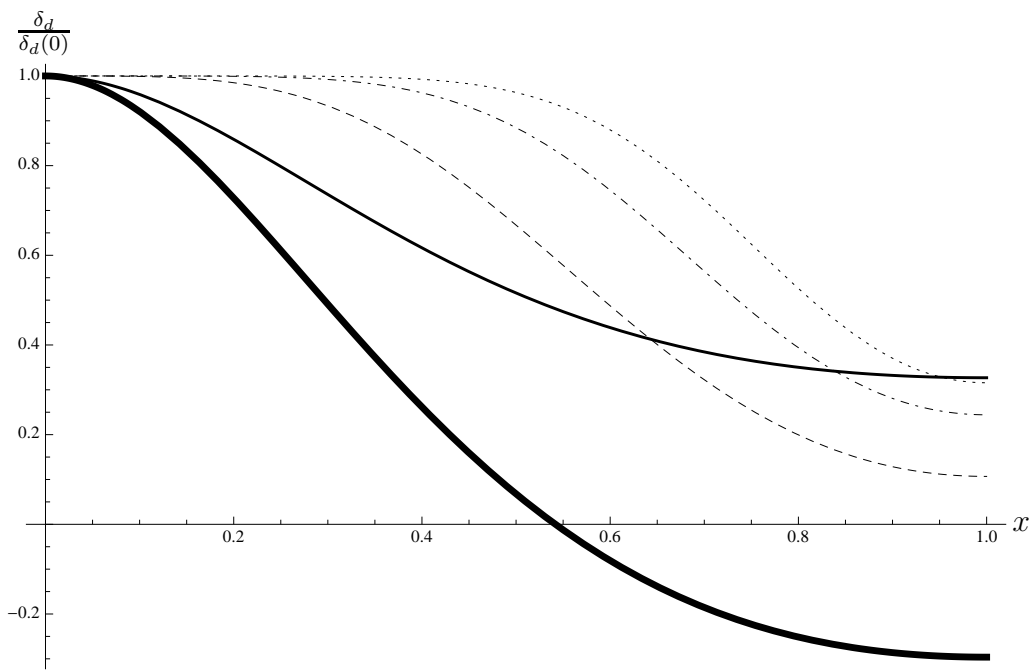}
	\caption{Plot of $\delta_d/\delta_d (x = 0)$ parametrizing the relative rate between de Sitter black hole and bubble production for $(d = 5, k = 1)$ (thick), $(d =5, k =2)$ (thin), $(d = 7, k = 1)$ (dashed), $(d = 9, k = 1)$ (dot-dashed) and $(d = 11, k = 1)$ (dotted) versus $x =  r_+/r_c$}
	\label{deltaplot}
	\end{figure}
	
	In five dimensions for $k = 1$ if $x$ is sufficiently large, specifically $x \gtrapprox 0.5418$,  bubble production dominates over black holes.  For higher dimensions or larger values of $k$, black hole production is always dominant.  Note the entire range of $x$ should not truly be trusted; black holes are only reliable semiclassical objects provided the horizon size is large compared to the Planck length, or if
\be
\frac{r_+}{l_p} = x \frac{r_c}{l_p} \gg 1
\ee	
Since it turns out that for the above instantons $r_c$ is comparable to $l$ (for $d = 5$, $0.655 \lessapprox \frac{r_c}{l} < 1$ and the allowed range of $\frac{r_c}{l}$ shrinks as the dimension increases), this means one can really only trust the black holes when
\be
x \gtrapprox \frac{l_p}{l}
\ee

Note then if the ratio $l/l_p$ is not very large, in five dimensions bubble production is often dominant in the region where the calculation is trustworthy.  On the other hand, if the ratio between the de Sitter length and Planck length is very large, the rates for both production processes are so small it is difficult to imagine any practical context where either one would be significant.  For example, in five dimensions if $l = 10 l_p$,
\be
P \sim e^{-(l/l_p)^3}  \approx 10^{-434}
\ee

\section{Discussion}

We have described an instanton describing the production of pairs of bubbles of nothing in odd higher dimensional de Sitter spaces.  While we do not known of similar solutions in even higher dimensional spaces, there is no obvious reason why they should not exist and may well be found in the future.  Given this process, if one has a higher dimensional theory with a cosmological constant comparable to the Planck scale, the resulting spacetime will not simply be the usual de Sitter (possibly with some black holes) but portions will have been removed by bubbles of nothing.
On the other hand, if the Planck scale is at least an order of magnitude smaller than the de Sitter length this process, as well as black hole production, is highly suppressed and the importance of the above work is mainly theoretical.   

The above pair production process provides a direct test for the proposition that cosmological horizons are associated with entropy and evade the objections a skeptic might make regarding black hole nucleation in de Sitter space.   This process also shows there is no topological obstruction to producing bubbles of nothing.   Noting that bubbles of nothing with the same topology as these de Sitter bubbles have been suggested as a possible generic instability of higher dimensions \cite{CopseyBubblesII}, this demonstrates generic spacetimes, including those which are asymptotically flat or asymptotically anti de Sitter, are not topologically safe from decay into such bubbles.  Hence, presuming one believes such instanton calculations (including those involving compact instantons) are reliable,  string theory, or any other consistent higher dimensional theory of quantum gravity, is forced to deal with such topologically nontrivial solutions.

\vskip .5cm
\centerline{\bf Acknowledgments}
\vskip .5cm

It is a pleasure to thank  S.Ross, S. Giddings, and D. Garfinkle for useful discussion and correspondence.  This work was supported by the Natural Sciences and Engineering Council of Canada.

 \appendix
 \setcounter{equation}{0}

\section{Sphere fibrations}
One may parametrize an odd dimensional round sphere by complex coordinates $Z^i$ such that $\Sigma_1 \vert Z^i \vert^2 =1$.  Then the metric on the sphere is given by
\be
d\Omega_{d - 2}^2 = \Sigma_{i} d Z^i d \bar{Z}^i = \Sigma_{i = 1}^{d-2} e_i^2
\ee
One may write a fiber as one of these one forms
\be
 (d \chi + A) = e_{d-2}
 \ee
and  the $CP^{N}$ metric as the sum of the remaining $e_i$
\be
d \Sigma^2 = \Sigma_{i = 1}^{d-3} e_i^2
\ee
Then the metric on the unit $(d-2)$-sphere may be written as
\be
d \Omega_{d-2} = (d \chi + A)^2 + d \Sigma^2
\ee
For specific explicit metrics we will use conventions where $\phi_i$, as well as $\chi$, have periods $2 \pi$ and $\theta_i$ have ranges $0 \leq \theta_0 \leq \pi/2$.
Specifically one may write $S^3$ using

\begin{eqnarray}
Z^1 &=& e^{i (\phi_1 + \chi)} \cos \theta_1\nonumber \\
Z^2 &= &e^{i \chi} \sin \theta_1 \nonumber 
\end{eqnarray}
and then one finds
\begin{eqnarray}
e_1 &=& d\theta_1 \nonumber  \\
e_2 &= &  \sin \theta_1  \, \cos \theta_1  \, d \phi_1 \nonumber  \\
e_3 &=&  d \chi + \cos^2 \theta_1 \, d\phi_1 \nonumber 
\end{eqnarray}
An $S_5$ may be written via
\begin{eqnarray}
Z^1 &=& e^{i (\phi_1 + \chi)} \cos \theta_1\nonumber  \\
Z^2 &= & e^{i (\phi_2 + \chi)} \sin \theta_1 \,  \cos \theta_2 \nonumber  \\
Z^3 &=& e^{i \chi} \sin \theta_1 \,  \sin \theta_2 \nonumber 
\end{eqnarray}
and then one finds
\begin{eqnarray}
e_1 &=& d\theta_1 \nonumber \\
e_2 &=& \sin \theta_1 \, d \theta_2 \nonumber  \\
e_3 &= &  \sin \theta_1  \, \cos \theta_1  \,( d \phi_1 - \cos^2 \theta_2 \, d \phi_2 ) \nonumber  \\
e_4 &=& \sin \theta_1 \, \sin \theta_2 \, \cos \theta_2 \, d \phi_2 \nonumber  \\
e_5 &=&  d \chi + \cos^2 \theta_1 \,  d\phi_1 + \sin^2 \theta_1 \,  \cos^2 \theta_2 \, d \phi_2 \nonumber 
\end{eqnarray}

$S_7$ may be written via
\begin{eqnarray}
Z^1 &=& e^{i (\phi_1 + \chi)} \cos \theta_1\nonumber  \\
Z^2 &= & e^{i (\phi_2 + \chi)} \sin \theta_1 \,  \cos \theta_2 \nonumber  \\
Z^3 &= & e^{i (\phi_3 + \chi)} \sin \theta_1 \,  \sin \theta_2 \, \cos \theta_3 \nonumber  \\
Z^4 &=& e^{i \chi} \sin \theta_1 \,  \sin \theta_2 \, \sin \theta_3 \nonumber 
\end{eqnarray}
and then one finds
\begin{eqnarray}
e_1 &=& d\theta_1 \nonumber  \\
e_2 &=& \sin \theta_1 \, d \theta_2 \nonumber  \\
e_3 &=& \sin \theta_1 \, \sin \theta_2 \, d \theta_3 \nonumber  \\
e_4 &= &  \sin \theta_1  \, \cos \theta_1  \,( d \phi_1 - \cos^2 \theta_2 \, d \phi_2 - \sin^2 \theta_2 \, \cos^2 \theta_3 \, d \phi_3 ) \nonumber  \\
e_5 &= &  \sin \theta_1  \, \sin \theta_2 \, \cos \theta_2  \,( d \phi_2 - \cos^2 \theta_3 \, d \phi_3 ) \nonumber  \\
e_6 &=& \sin \theta_1 \, \sin \theta_2 \, \sin \theta_3 \, \cos \theta_3 \, d \phi_3 \nonumber \\
e_7 &=&  d \chi + \cos^2 \theta_1 \,  d\phi_1 + \sin^2 \theta_1 \,  \cos^2 \theta_2 \, d \phi_2+ \sin^2 \theta_1 \,  \sin^2 \theta_2 \, \cos^2 \theta_3 \, d \phi_3 \nonumber 
\end{eqnarray}
and finally $S^9$ may be written using
\begin{eqnarray}
Z^1 &=& e^{i (\phi_1 + \chi)} \cos \theta_1\nonumber  \\
Z^2 &= & e^{i (\phi_2 + \chi)} \sin \theta_1 \,  \cos \theta_2 \nonumber  \\
Z^3 &= & e^{i (\phi_3 + \chi)} \sin \theta_1 \,  \sin \theta_2 \, \cos \theta_3 \nonumber  \\
Z^4 &= & e^{i (\phi_4 + \chi)} \sin \theta_1 \,  \sin \theta_2 \, \sin \theta_3 \, \cos \theta_4 \nonumber  \\
Z^5 &=& e^{i \chi} \sin \theta_1 \,  \sin \theta_2 \, \sin \theta_3 \, \sin \theta_4 \nonumber 
\end{eqnarray}
and then one finds
\begin{eqnarray}
e_1 &=& d\theta_1 \nonumber  \\
e_2 &=& \sin \theta_1 \, d \theta_2 \nonumber  \\
e_3 &=& \sin \theta_1 \, \sin \theta_2 \, d \theta_3 \nonumber  \\
e_4 &=& \sin \theta_1 \, \sin \theta_2 \, \sin \theta_3 \, d \theta_4 \nonumber  \\
e_5 &= &  \sin \theta_1  \, \cos \theta_1  \,( d \phi_1 - \cos^2 \theta_2 \, d \phi_2 - \sin^2 \theta_2 \, \cos^2 \theta_3 \, d \phi_3 - \sin^2 \theta_2 \, \sin^2 \theta_3 \, \cos^2 \theta_4 \, d \phi_4) \nonumber  \\
e_6 &= &  \sin \theta_1  \, \sin \theta_2 \, \cos \theta_2  \,( d \phi_2 - \cos^2 \theta_3 \, d \phi_3  - \sin^2 \theta_3 \, \cos^2 \theta_4 \, d \phi_4 ) \nonumber  \\
e_7 &= &  \sin \theta_1  \, \sin \theta_2 \, \sin \theta_3 \, \cos \theta_3  \,( d \phi_3 - \cos^2 \theta_4 \, d \phi_4 ) \nonumber  \\
e_8 &=& \sin \theta_1 \, \sin \theta_2 \, \sin \theta_3 \, \sin \theta_4 \cos \theta_4 \, d \phi_4 \nonumber  \\
e_9 &=&  d \chi + \cos^2 \theta_1 \,  d\phi_1 + \sin^2 \theta_1 \,  \cos^2 \theta_2 \, d \phi_2+ \sin^2 \theta_1 \,  \sin^2 \theta_2 \, \cos^2 \theta_3 \, d \phi_3 \nonumber \\
{} &{} &+ \sin^2 \theta_1 \,  \sin^2 \theta_2 \, \sin^2 \theta_3 \, \cos^2 \theta_4 \, d \phi_4  \nonumber
\end{eqnarray}


\begin{thebibliography}{99}
    
    \bibitem{EMpair}D. Garfinkle, S. B. Giddings, Phys. Lett. {\bf B256} (1991) 146;
 D. Garfinkle, S. B. Giddings and A. Strominger, Phys. Rev. {\bf D49} (1994) 958; 
H. F. Dowker, J. P. Gauntlett, D. A. Kastor and J. Traschen, Phys. Rev. {\bf D49} (1994) 2909;
H. F. Dowker, J. P. Gauntlett, S. B. Giddings and G. T. Horowitz, Phys. Rev. {\bf D50} (1994) 2662;
S.F. Ross, Phys. Rev. {\bf D51} (1995) 2813;  J. D. Brown, Phys. Rev. {\bf D51} (1995) 5725.

   
    \bibitem{RossMannBH} R. B. Mann and S. F. Ross, ``Cosmological production of charged black holes pairs,'' Phys. Rev. {\bf D52} (1995) 2254-2265  [arXiv:gr-qc/9504015]

\bibitem{cosmo} R. Bousso and S. W. Hawking, Phys. Rev. {\bf D52} (1995) 5659; R. Bousso and S. W. Hawking, Phys. Rev. {\bf D54} (1996) 6312; S. W. Hawking, Phys. Rev. {\bf D53} (1996) 3099; 
R. Garattini, Nucl. Phys. {\bf B57} (Proc. Suppl.) (1997) 316; R. Garattini, Nuovo Cimento  {\bf B113} 
(1998) 963; R. Garattini, Mod. Phys. Lett.  {\bf A13} (1998) 159; R. Garattini, Class. Quant. Grav.  {\bf 18} (2001) 571; M. Volkov and A. Wipf, Nucl. Phys.  {\bf B582} (2000) 313.

\bibitem{cstring} S. W. Hawking and S. F. Ross, Phys. Rev. Lett. {\bf 75} (1995) 3382;
 D. M. Eardley, G. T. Horowitz, D. A. Kastor and J.
Traschen, Phys. Rev. Lett. {\bf 75} (1995) 3390 ; H. F. Dowker and S. Thambyahpillai, Class. Quantum Grav. {\bf 20} (2003) 127; K. Hong and E. Teo, Class. Quantum Grav. {\bf 20} (2003) 3269.
 
\bibitem{dwall}  R. R. Caldwell, A. Chamblin and G. W. Gibbons, Phys. Rev. {\bf D53} (1996) 7103;
R. Mann, Class. Quantum Grav. {\bf 14} (1997) L109; R.B. Mann, Nucl. Phys. {\bf B516} (1998) 357; R. Bousso and A. Chamblin, Phys. Rev. {\bf D59} (1999) 084004.

\bibitem{combo} R. Emparan, Phys. Rev. Lett. {\bf 75} (1995) 3386; O. J. C. Dias and J. P. S. Lemos, Phys. Rev. {\bf D67} (2003) 064001; O. J. C. Dias and J. P. S. Lemos, Phys. Rev. {\bf D67} (2003) 084018; O. J. C. Dias and J. P. S. Lemos, Phys. Rev. {\bf D68} (2003) 104010; 
O. J. C. Dias and J. P. S. Lemos, Phys. Rev. {\ bf D69} (2004) 084006; O. J. C.
Dias, Phys. Rev. {\bf D70} (2004) 024007.

\bibitem{Ivan} I.S. Booth and R.B. Mann, Phys. Rev. Lett. {\bf 81} (1998) 5052; I.S. Booth and R.B. Mann, Nucl. Phys. {\bf B539} (1999) 267.

\bibitem{Ginsparg} P. Ginsparg and M. J. Perry, Nucl. Phys. {\bf B222} (1983) 245.

\bibitem{Bousso} R. Bousso, Rev. Mod. Phys. {\bf 74} (2002) 825.

 \bibitem{DiasLemos} O. J. C. Dias and J. P. S. Lemos, ``Pair creation of higher dimensional black holes on a de Sitter background,'' Phys. Rev. D. {\bf D70} (2004) 124023  [arXiv:hep-th/0410279]


       \bibitem{ClarksonMann1} R. Clarkson and R. B. Mann, ``Soliton Solutions to the Einstein Equations in Five Dimensions,'' Phys. Rev. Lett. {\bf 96} (2006) 051104   [arXiv:hep-th/0508109]
       
       \bibitem{CopseyBubblesII} K. Copsey, ``Bubbles Unbound II: AdS and the Single Bubble,'' JHEP {\bf 0710} (2007) 095   [arXiv:0706.3677]
       
        \bibitem{MannClarksonII} R. Clarkson and R. B. Mann, ``Eguchi-Hanson Solitons in Odd Dimensions,'' Class. Quant. Grav. {\bf 23} (2006) 1507-1524   [arXiv:hep-th/0508200]
       
          
          \bibitem{Wittenbubble}
 E.~Witten,
  ``Instability Of The Kaluza-Klein Vacuum,''
  Nucl.\ Phys.\ B {\bf 195} (1982) 481
    
      \bibitem{CSsol} G. Compere, K. Copsey, S. de Buyl, and R. B. Mann, ``Solitons in Five Dimensional Minimal Supergravity: Local Charge, Exotic Ergoregions, and Violations of the BPS Bound,'' JHEP {\bf 0912} (2009) 047 [arXiv: 0909.3289]
       
       
       \bibitem{GHdeSitter} G. W. Gibbons and S. W. Hawking, ``Cosmological event horizons, thermodynamics, and particle creation,'' Phys. Rev. {\bf D15} (1977) 2738
    
    \bibitem{GHbdy} G. W. Gibbons and S. W. Hawking, ``Action integrals and partition functions in quantum gravity,'' Phys. Rev. {\bf D15}  (1977) 2752.
    
    \bibitem{CdL} S. Coleman and F. De Luccia, ``Gravitaitonal effects on and of vacuum decay,'' Phys. Rev. {\bf D 21} (1980) 3305.
    
   \bibitem{conceptual}
T. Banks, ``Cosmological Breaking of Supersymmetry?  Or Little Lambda
goes back to the future 2,'' [arXiv:hep-th/0007146];
S. Hellerman, N. Kaloper and L. Susskind, ``String theory and
quintessence,'' JHEP {\bf 0106}, 003 (2001) [arXiv:hep-th/0104180];
W. Fischler, A. Kashani-Poor, R. McNees and S. Paban, ``The
acceleration of the Universe: A challenge for string theory,''
[arXiv:hep-th/0104181];
E. Witten, ``Quantum Gravity in de Sitter Space,'' [arXiv:hep-th/0106109];
A. Strominger, ``The dS/CFT Correspondence,'' JHEP {\bf 0110}, 034 (2001)
[hep-th/0106113].


\bibitem{MSS}
E. Silverstein, ``(A)dS Backgrounds from Asymmetric Orientifolds,''
[arXiv:hep-th/0106209]; A. Maloney, E. Silverstein and A. Strominger,
``de Sitter Space in Noncritical String Theory,'' [arXiv:hep-th/0205316].

\bibitem{Susskind}
L.~Dyson, J.~Lindesay and L.~Susskind,
``Is there really a de Sitter/CFT duality,''
JHEP {\bf 0208}, 045 (2002)
[arXiv:hep-th/0202163];
L.~Dyson, M.~Kleban and L.~Susskind,
``Disturbing implications of a cosmological constant,''
JHEP {\bf 0210}, 011 (2002)
[arXiv:hep-th/0208013];
N.~Goheer, M.~Kleban and L.~Susskind,
``The trouble with de Sitter space,''
arXiv:hep-th/0212209.

      
       
    \bibitem{RossHawkingDuality} S. W. Hawking and S. F. Ross, ``Duality between Electric and Magnetic Black Holes,'' ' Phys. Rev. {\bf D52} (1995) 5865-5876  [arXiv: hep-th/9504019]
    
  
    
    \bibitem{CopseyHorowitz1st} K. Copsey and G. T. Horowitz, ``The Role of Dipole Charges in Black Hole Thermodynamics,'' Phys. Rev. {\bf D73} (2006) 024015  [arXiv:hep-th/0505278]

  \end{thebibliography}
\end{document}